\documentclass[prl,amsfonts,amssymb,amsmath,floats,twocolumn,aps]{revtex4}

\usepackage[pdftex]{graphicx}
\usepackage{lipsum}
\usepackage{dcolumn}
\usepackage{bm}
\usepackage{color}
\usepackage{physics}
\usepackage[pdftex,colorlinks=true,linkcolor=blue,citecolor=blue,filecolor=blue]{hyperref}

\newcommand{\bi}{\begin{itemize}}
\newcommand{\ei}{\end{itemize}}
\newcommand{\be}{\begin{equation}}
\newcommand{\ee}{\end{equation}}
\newcommand{\ba}{\begin{eqnarray}}
\newcommand{\ea}{\end{eqnarray}}

\newcommand{\imag}{\mathrm{i}}
\begin{document} 

\title{Noise Correlations in time- and angular-resolved photoemission spectroscopy}

\author{Christopher Stahl and Martin Eckstein}
\affiliation{Department of Physics, University of Erlangen-Nuremberg, Staudtstra{\ss}e 7, 91058 Erlangen, Germany}
\begin{abstract}
In time-resolved photoemission experiments, more than one electron can be emitted from the solid by a single ultra-short pulse. We theoretically demonstrate how correlations between the momenta of  outgoing electrons relate to time-dependent two-particle correlations in the solid. This can extend the scope of time- and angular-resolved photoemission spectroscopy to probe superconducting and charge density fluctuations in systems without long-range order, and to reveal their dynamics independent of the electronic gap and thus unrestricted by the energy-time uncertainty. The proposal is  illustrated for superconductivity in a BCS model. An impulsive perturbation can quench the gap on ultrafast timescales, while  non-equilibrium pairing correlations persist much longer, even when electron-electron scattering beyond mean-field theory is taken into account. There is thus a clear distinction between a dephasing of the Cooper pairs and the thermalization into the normal state. While a measurement of the gap would be blind to such pairing correlations, they can be revealed by the angular correlations in photoemission.
\end{abstract} 
 
\maketitle 

ARPES (angular-resolved photoemission spectroscopy) is a powerful technique to probe the electronic structure in solids. With short laser-pulses in a pump-probe setup  one can moreover achieve femtosecond time-resolution, which has opened a unique path to explore the light-induced dynamics of collective phases in solids on ultra-short timescales \cite{Giannetti2016}. Time-resolved ARPES has been used to study ultra-fast quasiparticle-dynamics \cite{Smallwood2012, Rameau2016}, laser manipulation of electronic orders \cite{Schmitt2008, Rohwer2011,mor2017}, photo-induced Mott metal-insulator transitions \cite{Perfetti2006, wegkamp2014,Ligges2018}, and Floquet Bloch bands \cite{Wang2013}. An intriguing aim of the ultra-fast manipulation of condensed matter phases is to control orders like magnetism, charge density waves, or superconductivity. Although the corresponding order parameters are revealed  in the electronic spectra, e.g.~through the opening of a gap, some fundamental challenges remain to probe their dynamics using ARPES: (i) Spectroscopic probes are limited by the energy-time uncertainty, while the relevant dynamical processes in the destruction or formation of an order parameter $\phi$  (such as the superconducting condensate density) may be faster than the inverse of the gap $\Delta$ which identifies $\phi$ in the electronic spectrum \footnote{In this case, a subtle deconvolution of the spectrum from the probe envelope in time and frequency, or a suitable pulse shaping of the probe pulse may be needed to extract the relevant dynamical information \cite{Randi2017}.}, or happen on the same scale, as for the amplitude mode in superconductors \cite{Krull2014, Barankov2006,Kemper2015,Peronaci2015,Matsunaga2013}. (ii) The quantum state can exhibit strong fluctuations of the order parameter on the nanoscopic scale without forming a long-range order. Such non-equilibrium fluctuations may be an essential property of transient states in which new orders are stimulated by light \cite{Bauer2015, Ido2017, Lemonik2017, Dasari2018}, or, as we exemplarily discuss in this work, quenched on a short timescale. A measurement of the spectrum alone would be rather blind to this aspect of the ultrafast dynamics.

Modern time-of-flight detectors for ARPES image outgoing electrons with different momenta onto different pixels of a detector, and thus allow to simultaneously record two electrons which are emitted from a single ultra-short probe pulse into different angular directions. In this paper, we propose that information on time-dependent two-particle correlations in the solid can be revealed from the correlation between the emission into different directions, i.e., the shot-to-shot noise correlation on the detector. The intriguing potential in measuring noise has been demonstrated in various other settings. For example, noise correlation in time-of-flight measurements of the momentum distribution in ultra-cold gases \cite{Altman2004, Bloch2008} can distinguish different phases of the initially trapped quantum state, and the shot-to-shot variance of the reflectivity in optical pump-probe experiments has been used to detect squeezing of vibrational modes in a quartz crystal \cite{Esposito2015}, and to measure the current noise in photo-excited Bismuth to  probe non-thermal electrons \cite{Randi2017PRL3}.  

Two-particle correlations in photoemission have  been used previously to study electronic interactions in equilibrium \cite{Schuhmann2009,Pavlyukh2015,Truetschler2017}. The process discussed here is an emission of two electrons by two photons from the same ultrashort pulse (different from a double photo-emission where one photon leads to the emission of two electrons due to secondary processes), and thus allows a theoretical interpretation along the same lines as the conventional theory for time-resolved photoemission spectroscopy \cite{Freericks2009,Eckstein2008,Kemper2017}. We start from a Hamiltonian $H=H_{s}+H_{e}+H'$, where $H_s$ is the Hamiltonian of the solid, and $H_{e}=\sum_{p\sigma}E_{p}f^{\dagger}_{p\sigma}f_{p\sigma}^{\phantom{\dagger}}$ describes the emitted electrons with asymptotic momentum $p$ and spin $\sigma$ ($E_p=p^2/2m +W$, with the work function $W$). Note that $H_s$ may be time-dependent to incorporate the non-perturbative effect of a probe pulse, or other types of non-equilibrium perturbations. Electron emission is due to the coupling term
\begin{align}
H'=
\sum_{k,p,\sigma,\sigma'} S(t)e^{-\imag\Omega t} M_{k,p}^{\sigma,\sigma'} c^\dagger_{k\sigma}f_{p\sigma'}^{\phantom{\dagger}}+h.c.,
\end{align}
where 
$M_{k,p}^{\sigma,\sigma'} \equiv \delta_{\sigma,\sigma'} M_{k,p}$ are matrix elements (for notational simplicity we restrict the solid to one band with electron operators $c_{k\sigma}$), and $S(t)$ is the temporal envelope of the probe with frequency $\Omega$. The Hamiltonian $H$ has built in two basic assumptions which are commonly made in the theory of ARPES: (i) There is no interaction between electrons in outgoing states ($f_p$) and electrons in the solid ($c_k$), manifesting the sudden approximation. Furthermore, (ii), we neglect interactions between outgoing electron and space-charge effects, which is controlled by the excitation density. Finally, all illustrating calculations below are based on simple matrix elements  $M_{k,p}=M\delta_{k,p}$. This corresponds to full momentum conservation, as in two-dimensional materials where only the momentum parallel to the surface matters. Matrix element effects could easily be reinstated for an interpretation of real experiments.

A time-resolved ARPES measurement records the total population  $I^{(1)}_{p\sigma}=\langle n^f_{p\sigma}\rangle_{t=\infty}$ in an outgoing state ($n^f_{p\sigma}=f_{p\sigma}^\dagger f_{p\sigma}^{\phantom{\dagger}}$), which is accumulated over the entire probe pulse duration (until $t=\infty$). Because two electrons can be emitted by two photons from the same pulse, we can  measure the correlations $\Delta I_{p\sigma,p'\sigma'}=I^{(2)}_{p\sigma,p'\sigma'} - I^{(1)}_{p\sigma} I^{(1)}_{p'\sigma'}$ at $p\neq p'$, with $I^{(2)}_{p\sigma,p'\sigma'}=\langle n^f_{p\sigma}n^f_{p'\sigma'}\rangle_{t=\infty}$.  For a weak probe pulse, all signals are obtained using the leading-order time-dependent perturbation theory  in the coupling $H'$, which is $2$nd order for $I^{(1)}$ and $4th$ order for $I^{(2)}$. We assume that at $t=-\infty$ the outgoing states are empty, and the solid is described by its initial density matrix $\rho_0^s$. Switching to interaction representation in $H'$ yields $I^{(2)}_{p\sigma,p'\sigma'}=\langle  \mathcal{S}^\dagger  n^f_{p\sigma}n^f_{p'\sigma'} \mathcal{S}\rangle_0$, where $ \mathcal{S}=T_t e^{-\imag\int_{-\infty}^{+\infty} d\bar t H'(\bar t)}$ is the S-matrix, and $\langle\cdots\rangle_0$ the initial state expectation value. Because the initial state does not contain outgoing electrons, the leading order expansion of $\mathcal{S}$ and $\mathcal{S}^\dagger$ in terms of $H'$ is second order and must contain both $f_{p\sigma}^\dagger$ and  $f_{p'\sigma'}^\dagger$ (both $f_{p\sigma}$ and  $f_{p'\sigma'}$) in $\mathcal{S}$ ($\mathcal{S}^\dagger$), respectively. After the expansion, the expectation value factorizes for the solid and the outgoing states, so that the result can be expressed in terms of one- and two-point Green's functions of the solid, $G(1,1')=\langle c(1)^\dagger  c(1')\rangle_0$ and 
\begin{align}
\label{nhebnnwnwx}
G(1,2,2',1')
=
&
\langle
T_{\bar t}[c(1)^\dagger c(2)^\dagger ]
T_{t} [c(2') c(1')]
\rangle_0.
\end{align}
Here $1\equiv (k_1,\sigma_1,t_1)$ etc., is short for space-time variables, and $T_{t}$ ($T_{\bar t}$) is the (anti)-time ordering operator for Fermions. Finally all terms can be combined to (see Appendix)
\begin{align}
I^{(1)}_{p\sigma}
&=
\int \!\!d1 d1'
M^{p,\sigma}_{1,1'}
G(1,1'),
\label{,efewnjqebckc}
\\
I^{(2)}_{p\sigma,p'\sigma'}
&=
\int \!\!d1 d1' d2 d2'\,
M^{p,\sigma}_{1,1'}
M^{p',\sigma'}_{2,2'}
G(1,2,2',1'),
\label{,efewnjqebckc888}
\end{align}
where $\int d1=\sum_{k_1,\sigma_1}\int _{-\infty}^\infty dt_1$, and 
\begin{align}
M^{p,\sigma}_{1,1'}
=
M_{k_1,p}^{\sigma_1,\sigma}
(M_{k_1',p}^{\sigma_1',\sigma})^*
S(t_1)
S(t_1')^*
e^{\imag(E_p-\Omega)(t_1-t_1')}.
\end{align}
The expression for $I^{(1)}_{p\sigma}$ is the conventional expression for time-resolved ARPES \cite{Freericks2009, Eckstein2008}, 
which can be understood as a time-dependent filter $M(t,t')$ applied to the single-particle propagator \cite{Randi2017}. Equation~\eqref{,efewnjqebckc888} provides an analogous view on two-particle quantities. 

To illustrate the use of noise correlation in time-resolved ARPES, one can consider an ideal ultra-short pulse $S(t)=A\delta(t-t_0)$. In this case, Eqs.~\eqref{,efewnjqebckc} and \eqref{,efewnjqebckc888} yield $I^{(1)}_{p\sigma} = |AM|^2 \langle n^c_{p\sigma}\rangle_{t=t_0}$ and 
\begin{align}
\label{,mxqnsk}
\Delta I^{(2)}_{p\sigma,p'\sigma'} = |AM|^4 
\big(\langle n^c_{p\sigma} n^c_{p'\sigma'}\rangle
-
\langle n^c_{p\sigma}\rangle
\langle n^c_{p'\sigma'}\rangle
\big)_{t=t_0},
\end{align}
where $n^c_{p\sigma}=c_{p\sigma}^\dagger c_{p\sigma}$ is the momentum occupation, and $\langle \mathcal{O}\rangle_{t} $ is the expectation value of an operator {\em in the solid} at time $t$. The angular correlations thus directly yield the momentum correlations in the solid, which can provide unique information on the state. In the BCS wave function, e.g.,  $\langle n^c_{k\uparrow}n^c_{-k\downarrow} \rangle -\langle n^c_{k\uparrow}\rangle\langle n^c_{-k\downarrow} \rangle= |\langle c_{k\uparrow} c_{-k\downarrow}\rangle |^2$ is a direct measure of pairing correlations, while $\langle n^c_{k\sigma}\rangle$ remains smooth throughout the superconducting transition.

By looking at different pairs $k,k'$ ($k\neq k'$) in $\Delta I_{k,k'}$, different symmetry broken phases can be characterized (charge-density waves, superconductivity, etc.). In the following we provide an illustrative example for using the noise correlations in the study of  superconductivity. We start the discussion from the Hubbard model
\begin{equation}
H=-J\sum_{\langle i,j \rangle,\sigma}c^\dagger_{i\sigma}c_{j\sigma}+U/2\sum_{i,\sigma}n_{i\sigma}n_{i-\sigma},
\label{eqn:Hubbard}
\end{equation}
which is the paradigmatic Hamiltonian to describe the physics of interacting electrons. Here $J$ is a hopping between nearest neighbor sites on a lattice, and $U$ is an on-site interaction. We choose an attractive interaction $U<0$,
which leads to s-wave superconductivity. To understand the rich non-equilibrium dynamics in superconductors, it is illustrative to recapitulate first the time-dependent mean-field solution. By decoupling the interaction term in the Cooper channel, the BCS-Hamiltonian $H_{BCS}=\sum_k\hat \psi^\dagger_k\hat h_k\hat \psi_k$ is obtained. Here  $\hat \psi_{k}=(c_{k\uparrow},\, c^\dagger_{-k\downarrow})^T$ is the Nambu-Spinor and $\hat h_k=\hat \sigma_z\epsilon_k+\hat \sigma_x\Delta'-\hat \sigma_y\Delta''$, with the electron dispersion $\epsilon_k$ and the gap  $\Delta=\Delta'+\imag\Delta''=U\sum_{k}\expval{c_{-k\downarrow} c_{k\uparrow}}$. The  BCS-Hamiltonian can be written in terms of the Anderson pseudo-spins $\vec{s}_k=\tfrac12\hat \psi_k^\dagger\vec{\sigma}\hat \psi_k$ \cite{Anderson1958}, which follow the equation of motion $\dot{\vec{s}}_k=\vec{B}_k\cross \vec{s}_k$ with the pseudo magnetic field $\vec{B}_k=(2\Delta',-2\Delta'', 2\epsilon_k)$. This defines an integrable set of coupled linear differential equations with an infinite number of conserved quantities \cite{Barankov2006,Dzero2006,Altshuler2006}.  A simple protocol such as a sudden quench or ramp of the interaction can lead to collective amplitude modes or an ultra-fast vanishing of the gap \cite{Barankov2006,Krull2014}. It must be emphasized that the mean-field-dynamics is highly non-thermal, even after a melting of the gap. For example, after a quench of the interaction to $U=0$ in $H_{BCS}$, the gap exponentially decays like $\Delta(t) \sim e^{-2t\Delta(0)}$, while the Cooper-pair correlations $F_k=|\langle c_{-k\downarrow}c_{k\uparrow}\rangle|$ at each $k$ remain nonzero, because the Anderson pseudo-spins $\vec{s}_k$ simply precess at different frequencies, such that the global order only dephases \cite{Dzero2006}. In contrast, thermalization to a normal state above $T_c$ (e.g., due to electron-electron scattering) would imply $F_k=0$. 

In the following we demonstrate that thermalization and dephasing of superconducting order can be distinct even when realistic electron-electron scattering beyond mean-field theory is taken into account, and that  the noise correlations provide a unique measure to distinguish them experimentally.
We examine a simple quench or ramp of the interaction, which initiates dynamics representative for a generic impulsive excitation:
The final value of $U$ determines  the electron-electron scattering and the pairing interaction during the dynamics, while the quench or ramp amplitude mainly sets  the excitation density. To incorporate electron-electron scattering beyond mean-field theory, the Hubbard model is solved using non-equilibrium dynamical mean-field theory (DMFT)  and an impurity solver based on iterated perturbation theory \cite{aoki2014_rev}.  We use a semi-elliptic density of states $D(\epsilon)$, where the half-bandwidth $W=2$ sets the unit of energy and time ($\hbar=1$). DMFT  gives access to all normal and anomalous single-particle Green's function in the lattice, in particular the condensate density $\phi=\sum_{k}\expval{c_{-k\downarrow} c_{k\uparrow}}$  and the individual $F_k$.
Momentum dependent quantities $g_k$ are represented as functions of the band energy $\epsilon_k\in(-2,2)$ (Fermi energy $\epsilon_F=0$), and  momentum averages are given by the integral $\sum_kg_k\equiv\int d\epsilon D(\epsilon) g(\epsilon)$.
 
\begin{figure}
\includegraphics[scale=1]{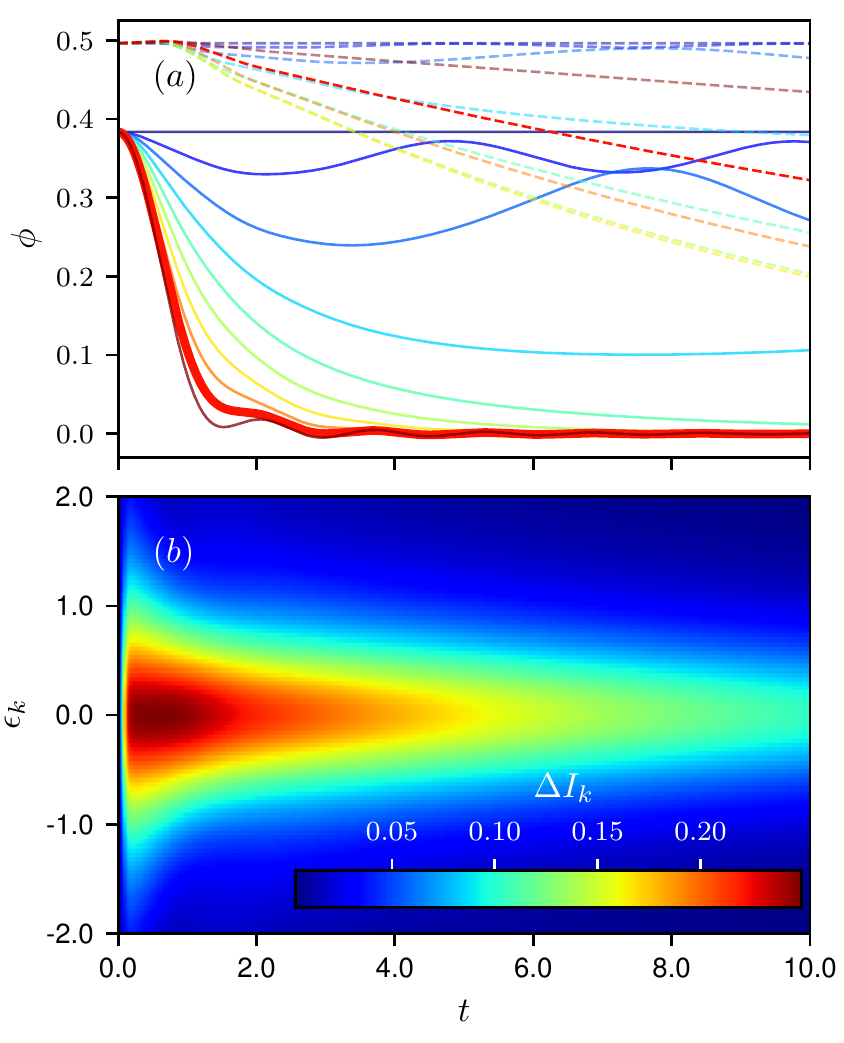}
\caption{a) Order parameter $\phi$ (\textit{solid}) and pairing correlations $F_{k_f}$ (\textit{dashed}) at the Fermi surface ($\epsilon_k=0$) after an interaction quench with $\Delta U=0.0,0.3,\ldots,2.7$ (top to bottom). The bold line correspond to $\Delta U =2.4$. 
b) Numerical simulation of the noise correlation measurement for $\Delta U =2.4$: $\Delta I_k(t)$, as obtained from Eq.~\eqref{eqn:BCS-Variance} with a short pulse $S(\tau)=\sqrt{100/\pi}e^{-100(\tau-t)^2}$ centered around time $t$.}
\label{fig:Quenched}
\end{figure}

Figure~\ref{fig:Quenched}a shows the order parameter $\phi(t)$ (solid lines) and the Cooper pair amplitude $F_{k_f}$ (dashed lines) at the Fermi-surface, after a sudden quench of the interaction from an initial value $U_0=-3$ to $U=U_0+\Delta U$. For weak excitations $\Delta U$,  the order parameter and $F_{k_f}$ oscillate  with a small amplitude, while both decays to zero  for large $\Delta U$ (e.g., dark red lines). In general,  $F_{k_f}$ decays much slower than $\phi$, even at relatively large interactions {\em and} excitations strong enough to melt the gap (see, e.g., the bold curve for $U=-0.6$). Hence there is a large time window where the vanishing of the order parameter $\phi$ is mainly due to dephasing, in spite of electron-electron scattering. 
This behavior can be unravelled by the noise correlation measurement. As the condensate of Cooper pairs is formed by electrons with opposite momentum $p$ and $-p$, it is natural to measure the correlations $\Delta I_p \equiv \tfrac12\sum_{\sigma,\sigma'}\Delta I_{p\sigma,-p\sigma'}$. For the  BCS-Hamiltonian one could use Wick's theorem to decouple the two-point function \eqref{nhebnnwnwx}.  The only nonvanishing contribution to the connected Green's function $G(1,2,2',1')-G(1,1')G(2,2')$ which enters the fluctuations $\Delta I_p$  is therefore related to an anomalous Green's function $\bar G_p(t,t')=\langle T_{\bar{t}}\,[c^\dagger_{p\uparrow}(t')c^\dagger_{-p\downarrow}(t)]\rangle$ [c.f.~Eqs.~\eqref{,efewnjqebckc888} and \eqref{,efewnjqebckc}],
\begin{equation}
\Delta I_p
=\!\abs{\int\! \dd t'\dd t''S(t')S(t'') \bar G_p(t',t'')e^{\imag(E_p-\Omega)(t'+t'')}}^2,
\label{eqn:BCS-Variance}
\end{equation}
where we set $M_{k,p}^{\sigma_1,\sigma}=\delta_{k\sigma_1,p\sigma}$ as explained above. As long as the system is initially deeply in the symmetry broken phase, these anomalous terms capture the leading contribution to the two-particle Green's function even beyond mean-field theory (apart from vertex corrections).
We therefore simply evaluate Eq.~\eqref{eqn:BCS-Variance} using the DMFT solution. The  simulated ARPES noise correlations, shown in the  Fig.~\ref{fig:Quenched}b, directly reveal the presence of Cooper pair correlations beyond the vanishing of $\phi$. A complementary tr-ARPES can detect the vanishing of $\phi$ by the closing of the spectral gap, so that the dephasing of the superconducting state can be identified. Vertex corrections to Eq.~\eqref{eqn:BCS-Variance} would complicate a quantitative prediction of the value of $\Delta I_k$, but the very different timescales for the two and one-particle dynamics should remain a clear signature for experiment.

\begin{figure}
\includegraphics[scale=1]{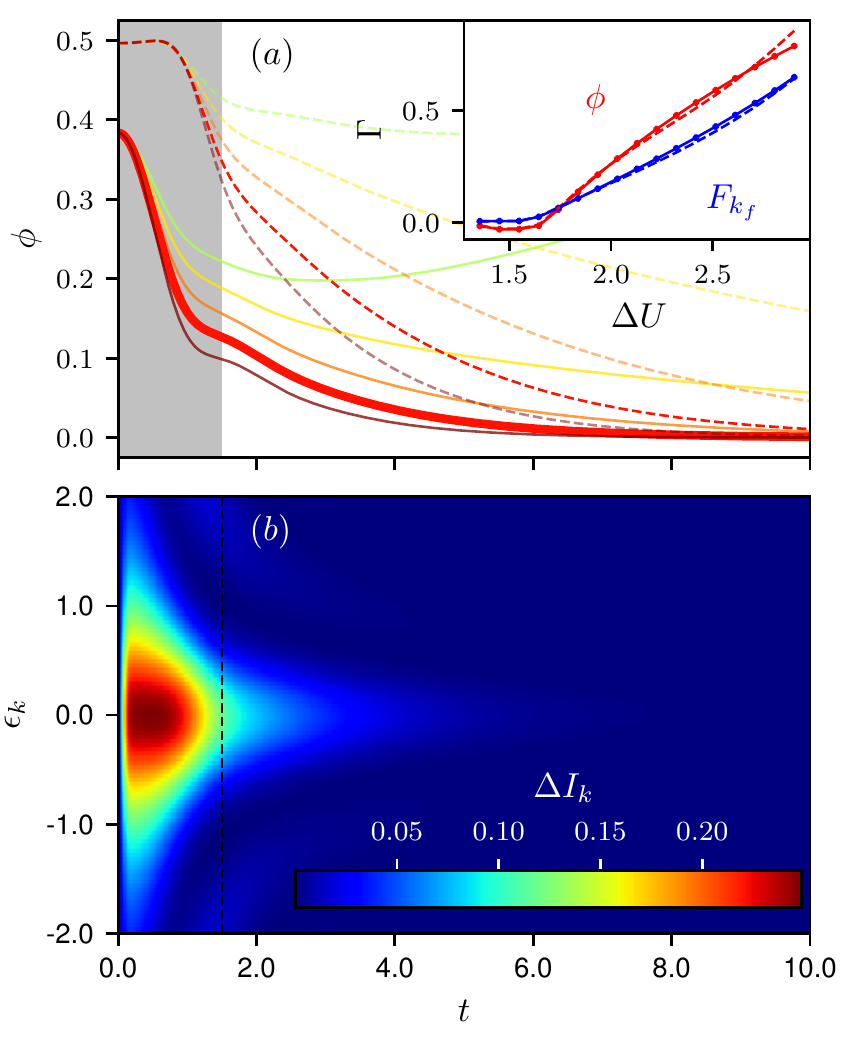}
\caption{a) Order parameter $\phi$ (solid lines) and Cooper pair correlations $F_{k_f}$ (dashed lines) for the ramp protocol with different excitation densities $\Delta U=1.5,1.8,\ldots,2.7$ (from top to bottom). The bold line shows $\Delta U =2.4$. The shaded area highlights ramp duration period. \textit{Inset:} Decay rate $\Gamma$ obtained by an exponential fit to $\phi(t)$ (solid red), the momentum averaged noise $\sum_k \sqrt{\Delta I_{k}(t)}$ (dashed red), $F_{k_f}$ (solid, blue) and $\sqrt{\Delta I_{k_F}(t)}$ (dashed  blue) against $\Delta U$. b) Numerical simulation of the noise correlation measurement $\Delta I_{k}(t)$ for energies around the Fermi edge for $\Delta U =2.4$. }
\label{fig:Ramped}
\end{figure}

In the quench protocol, a strong excitation of the system simultaneously implies weak final interactions $U$. To simulate a strong  impulsive excitation at large $U$, we perform a short pulse-shaped ramp of the interaction of duration $\tau$, $U(t)=U_0+\Delta U/2\theta(\tau-t) (1+\cos(\pi t/\tau))$. Figure~\ref{fig:Ramped}a shows the resulting $\phi(t)$ (solid lines) and $F_{k_f}$. The relaxation  dynamics is analyzed in a regime of relatively large $U=-3$, which is close to the maximum  of the transition temperature $T_c(U)$ in the phase diagram, corresponding to the crossover into the strong-coupling regime of a BEC of preformed pairs. Similar to Fig~\ref{fig:Quenched}a, amplitude mode oscillations or a melting of the gap are observed depending on the excitation $\Delta U$, but the larger electron-electron scattering now leads to a rapid relaxation of both the order parameter $\phi$ and $F_{k_f}$. The decay rates $\Gamma_\phi$ and $\Gamma_{F}$ of $\phi$ and of $F_{k_f}$ are of the same order, as shown by the solid lines in the inset of Fig.~\ref{fig:Ramped}a. (Both $\Gamma_\phi$ and $\Gamma_{F}$ show a slow-down at the threshold $\Delta U\approx 1.6$ for the melting of the order.) The noise correlations (Fig.~\ref{fig:Ramped}b), allow to probe the dynamics of these quantities. In particular, by fitting an exponential decay $\exp(-\Gamma t)$ to the simulated data for $\sqrt{\Delta I_{k}(t)}$ and the momentum average $\sum_k \sqrt{\Delta I_{k}(t)}$ one can closely recover the corresponding rates $\Gamma_\phi$ and $\Gamma_F$ (inset).   In the present case, pairing interactions and scattering are controlled by the same microscopic interaction $U$, so that the melting happens on timescales still larger than $\hbar/\Delta$, which could be resolved in tr-ARPES. In general, however, there is no fundamental limitation for how fast $\phi$ can be quenched to zero, and the noise correlation measurement, which is independent of the spectral information, grants access to the pair correlations on time scales beyond the energy-time uncertainty limitations of tr-ARPES.

In conclusion, we have proposed to use the angular correlations in ARPES to characterize time-dependent two-particle correlations in the solid. The latter can be expected to dominate non-equilibrium states, but are hard to measure with other techniques. Exemplarily, we showed that the dephasing of the individual order parameter fluctuations can dominate the fast decay of superconductivity, even when electron-electron scattering beyond mean-field  theory is included. Beyond this example, the noise correlation measurement  could be used to probe transient charge-density wave fluctuations (with correlations between momenta that differ by the nesting vector), excitonic correlations, or magnetic order (to provide a different view on intriguing phenomena such as the melting of magnetic order \cite{Eiche2017}), and possibly  help to reveal fundamental phenomena such as a possible non-thermal criticality in solids \cite{Berges2008,Tsuji2013}. In future work, it will also be interesting to monitor transient states which can have enhanced correlations but no long-range. Furthermore, while the example of this paper uses a short probe pulses to reveal more or less instantaneous correlations, the general result Eq.~\eqref{,efewnjqebckc888} shows that even {\em dynamical} time and energy-dependent two-particle quantities can be extracted. Another interesting perspective is the measurement of noise correlations during the application of a pulse, e.g., to reveal time-dependent correlations in Floquet driven states. This requires a gauge invariant reformulation of the theory in the presence of an external vector potential, analogous to standard photoemission \cite{Kemper2017}. In general, we conclude that the measurement of noise correlations in ARPES, though technically challenging, may give unique access to two-particle correlations in solids, which provides information that is indispensable to characterize the spatio-temporal evolution of non-equilibrium states.\\

\begin{acknowledgements}
We thank U. Bovensiepen and D. Fausti for useful discussions. We acknowledge the financial support from the DFG Project 310335100, and the ERC starting grant No.~716648.
\end{acknowledgements}

\appendix
\section{Appendix: Noise correlation Derivation}
In this section we derive the general result for noise correlation $\Delta I_{p\sigma,p'\sigma'}$ from fourth order perturbation theory as explained in the main text. We start with the expression $I^{(2)}_{p\sigma,p'\sigma'}=\expval{\mathcal{S}^\dagger n_{p\sigma}^fn_{p'\sigma'}^f\mathcal{S}}_0$, where 
\begin{align}
\mathcal{S}=T_te^{-\imag\int_{-\infty}^\infty\dd \bar{t}H'(\bar{t})}
\end{align} 
is the S-matrix in interaction representation with respect to $H'$. Because initial state contain no electrons in the $f$-states, the only contribution in fourth order will be
\begin{widetext}
\begin{align}
\nonumber
&\frac{1}{4}\expval{\int_{-\infty}^\infty\dd t_1\int_{-\infty}^\infty\dd t_2 T_{\bar{t}}[H'(t_1)H'(t_2)]n_{p\sigma}^f n_{p'\sigma'}^f\int_{-\infty}^\infty \dd t_{1'}\int_{-\infty}^\infty\dd t_{2'} T_t[H'(t_{1'})H'(t_{2'})]}_0\\
=&\sum_{\substack{k_1,k_2,k_{1'},k_{2'}\\p_1,p_2,p_{1'},p_{2'}}}\sum_{\substack{\sigma_1,\sigma_2,\sigma_{1'},\sigma_{2'}\\\tau_1,\tau_2,\tau_{1'},\tau_{2'}}}\int_{-\infty}^\infty\dd t_1\int_{-\infty}^{t_1} \dd t_2\int_{-\infty}^{\infty}\dd t_{1'}\int_{-\infty}^{t_{1'}} \dd t_{2'}S(t_1)^\ast S(t_2)^\ast S(t_{1'})S(t_{2'})e^{-\imag\Omega(t_{1'}+t_{2'}-t_1-t_2)}
\nonumber
\\
&\times
(M_{k_1,p_1}^{\sigma_1,\tau_1})^\ast⁽M_{k_2,p_2}^{\sigma_2,\tau_2})^\ast M_{k_{2'},p_{2'}}^{\sigma_{2'},\tau_{2'}}M_{k_{1'},p_{1'}}^{\sigma_{1'},\tau_{1'}}\expval{c^\dagger(2) c^\dagger(1)c(1')c(2')}_0^c\otimes\expval{f(\bar{2})f(\bar{1})f_{p\sigma}^{\dagger}f_{p\sigma} f_{p'\sigma'}^{\dagger}f_{p'\sigma'}f^{\dagger}(\bar{1}')f^{\dagger}(\bar{2}')}^f_0,
\label{,macbcb01}
\end{align}
where we introduced the super-indices $i=(k_i,\sigma_i,t_i)$ and $\bar{i}=(p_i,\tau_i,t_i)$.  
Due to the restriction of free electrons and the assumption that $\ket{\psi^f}$ is empty, we can evaluate the expectation value over $f$ using Wick's theorem and obtain four terms for $p\neq p'$:
\begin{align}
&\expval{f(\bar{2})f_{p\sigma}^{\dagger}}
\expval{f(\bar{1})f_{p'\sigma'}^{\dagger}}
\expval{f_{p\sigma}f^{\dagger}(\bar{2}')}
\expval{f_{p'\sigma'}c^{\dagger}(\bar{1}')}
+
\expval{f(\bar{2})f_{p'\sigma'}^{\dagger}}
\expval{f(\bar{1})f_{p\sigma}^{\dagger}}
\expval{f_{p\sigma}f^{\dagger}(\bar{1}')}
\expval{f_{p'\sigma'}f^{\dagger}(\bar{2}')}\nonumber
\\
&-\expval{f(\bar{2})f_{p\sigma}^{\dagger}}
\expval{f(\bar{1})f_{p'\sigma'}^{\dagger}}
\expval{f_{p\sigma}f^{\dagger}(\bar{1}')}
\expval{f_{p'\sigma'}f^{\dagger}(\bar{2}')}
-
\expval{f(\bar{2})f_{p'\sigma'}^{\dagger}}
\expval{f(\bar{1})f_{p\sigma}^{\dagger}}
\expval{f_{p\sigma}f^{\dagger}(\bar{2}')}
\expval{f_{p'\sigma'}f^{\dagger}(\bar{1}')}.
\label{,macbcb}
\end{align}
The expectation value $\expval{c^f(\bar{i})c^{f\dagger}(\bar{j})}$ is given by $\delta_{\tau_i,\tau_j}\delta_{p_i,p_j} e^{-\imag E_{p_{i}}(t_i-t_j)}$. Equation \eqref{,macbcb} can be reinserted into Eq.~\eqref{,macbcb01}. After summing over $\tau_1, \tau_{1'}, \tau_2,\tau_{2'}, p_1, p_{1'}, p_2, p_{2'}$ we obtain:
\begin{align}
\sum_{\substack{k_1,k_2,k_{1'},k_{2'}\\\sigma_1,\sigma_2,\sigma_{1'},\sigma_{2'}}}
&\int_{-\infty}^\infty\dd t_1\int_{-\infty}^{t_1} \dd t_2\int_{-\infty}^{\infty}\dd t_{1'}\int_{-\infty}^{t_{1'}} \dd t_{2'}
S(t_1)^\ast S(t_2)^\ast S(t_{1'})S(t_{2'})
e^{-\imag\Omega(t_{1'}+t_{2'}-t_1-t_2)}\nonumber\\
\times&\Big[
(M_{k_1,p'}^{\sigma_1,\sigma'})^\ast
(M_{k_2,p}^{\sigma_2,\sigma})^\ast
M_{k_{2'},p}^{\sigma_{2'},\sigma}
M_{k_{1'},p'}^{\sigma_{1'},\sigma'}\expval{c^\dagger(2) c^\dagger(1)c(1')c(2')}_0^c e^{-\imag E_p(t_2-{t_{2'}})}e^{-\imag E_{p'}(t_1-t_{1'})}\nonumber
\\
&+
(M_{k_1,p}^{\sigma_1,\sigma})^\ast
(M_{k_2,p'}^{\sigma_2,\sigma'})^\ast
M_{k_{2'},p'}^{\sigma_{2'},\sigma'}
M_{k_{1'},p}^{\sigma_{1'},\sigma}\expval{c^\dagger(2) c^\dagger(1)c(1')c(2')}_0^ce^{-\imag E_p(t_1-{t_{1'}})}e^{-\imag E_{p'}(t_2-t_{2'})}
\\
&-(M_{k_1,p'}^{\sigma_1,\sigma'})^\ast
(M_{k_2,p}^{\sigma_2,\sigma})^\ast
M_{k_{2'},p'}^{\sigma_{2'},\sigma'}
M_{k_{1'},p}^{\sigma_{1'},\sigma}
\expval{c^\dagger(2) c^\dagger(1)c(1')c(2')}_0^c e^{-\imag E_p(t_2-{t_{1'}})}e^{-\imag E_{p'}(t_1-t_{2'})}\nonumber
\\
&-(M_{k_1,p}^{\sigma_1,\sigma})^\ast
(M_{k_2,p'}^{\sigma_2,\sigma'})^\ast
M_{k_{2'},p}^{\sigma_{2'},\sigma}
M_{k_{1'},p'}^{\sigma_{1'},\sigma'}
\expval{c^\dagger(2) c^\dagger(1)c(1')c(2')}_0^c e^{-\imag E_p(t_1-{t_{2'}})}e^{-\imag E_{p'}(t_2-t_{1'})}\Big].\nonumber
\end{align}
By relabeling the indices in the second ($(1,{1'})\leftrightarrow(2,{2'})$), third ($(1')\leftrightarrow(2')$), and fourth ($1 \leftrightarrow 2$) term of the integrand, one can rewrite the expression in one integral by reintroducing the time ordering operators and obtain the result given in the main text:
\begin{align}
I^{(2)}_{p\sigma,p'\sigma'}=\sum_{\substack{k_1,k_2,k_{1'},k_{2'}\\\sigma_1,\sigma_2,\sigma_{1'},\sigma_{2'}}}
&\int_{-\infty}^\infty\dd t_1\int_{-\infty}^{\infty} \dd t_2\int_{-\infty}^{\infty}\dd t_{1'}\int_{-\infty}^{\infty} \dd t_{2'}
S(t_1)^\ast S(t_2)^\ast S(t_{1'})S(t_{2'})e^{-\imag\Omega(t_{1'}+t_{2'}-t_1-t_2)}\nonumber\\
&\times 
(M_{k_1,p'}^{\sigma_1,\sigma'})^\ast
(M_{k_2,p}^{\sigma_2,\sigma})^\ast
M_{k_{2'},p}^{\sigma_{2'},\sigma}
M_{k_{1'},p'}^{\sigma_{1'},\sigma'}\expval{T_{\bar{t}}[c^\dagger(2) c^\dagger(1)]T_{t}[c(1')c(2')]}_0^c e^{-\imag E_p(t_2-{t_{2'}})}e^{-\imag E_{p'}(t_1-t_{1'})}.
\end{align}
For the contribution $I^{(1)}_{p,\sigma}$ we follow the same route, but only need to go to a second order expansion as $\expval{n^{f}_{p\sigma}}_{0}=0$ and therefore the only contribution to $I^{(1)}_{p,\sigma}I^{(1)}_{p',\sigma'}$ in fourth order comes from the second order expansion:
\begin{align}
I^{(1)}_{p,\sigma}=\sum_{\substack{k_1,k_{1'},p_1,p_{1'}\\ \sigma_1,\sigma_{1'},\tau_1,\tau_{1'}}}\int_{-\infty}^{\infty}\dd t_1\int_{-\infty}^{\infty} \dd t_{1'} S(t_1)^\ast S(t_{1'})e^{-\imag\Omega(t_{1'}-t_1)}
(M_{k_1,p_1}^{\sigma_1,\tau_1})^\ast M_{k_{1'},p_{1'}}^{\sigma_{1'},\tau_{1'}}\expval{c^\dagger(1) c(1')}_0^c\otimes\expval{f(1)f^\dagger_{p, \sigma} f_{p,\sigma} f^\dagger(1')}^f_0,
\end{align}
where we use the same index convention as before. Again a expansion of the expectation value over $f$ using Wick's theorem yields a term $\expval{f(1)f^\dagger_{p,\sigma}}\expval{f_{p,\sigma}f^\dagger(1')}$, which can be evaluated to $\delta_{p_1,p}\delta_{p_{1'},p}\delta_{\tau_1,\sigma}\delta_{\tau_{1'},\sigma}e^{\imag E_{p}(t_{1}-t_{1'})}$. After reinserting the Kronecker-deltas and contracting the sums, we arrive at the final expression of the main text:
\begin{align}
I^{(1)}_{p,\sigma}=\sum_{k_1,k_{1'},\sigma_1,\sigma_{1'}}\int_{-\infty}^{\infty}\dd t_{1}\int_{-\infty}^{\infty}\dd t_{1'}
S(t_1)^\ast S(t_{1'})e^{-\imag\Omega(t_{1'}-t_1)} (M_{k_1,p}^{\sigma_1,\sigma})^\ast M_{k_{1'},p}^{\sigma_{1'},\sigma}\expval{c^\dagger(1) c(1')}_0^ce^{-\imag E_{p}(t_{1}-t_{1'})}.
\end{align}
\end{widetext}
As explained in the main text the noise correlation are given by $\Delta I_{p\sigma,p',\sigma'}=I^{(2)}_{p\sigma,p'\sigma'}-I^{(1)}_{p,\sigma}I^{(1)}_{p',\sigma'}$.

\bibliographystyle{apsrev4-1}

%

\end{document}